\documentclass[twocolumn,aps]{revtex4}

\pdfpageattr {/Group << /S /Transparency /I true /CS /DeviceRGB>>}

\usepackage{graphicx} 
\usepackage{overpic}

\newcommand{\ket}[1]{\left|#1\right\rangle}

\begin{document}

\title{Accurate simulations of planar topological codes cannot use cyclic boundaries}

\author{Austin G. Fowler}

\affiliation{Centre for Quantum Computation and Communication
Technology, School of Physics, The University of Melbourne, Victoria
3010, Australia}

\date{\today}

\begin{abstract}
Cyclic boundaries are used in many branches of physics and mathematics, typically to assist the approximation of a large space. We show that when determining the performance of planar, fault-tolerant, topological quantum error correction, using cyclic boundaries leads to a significant underestimate of the logical error rate. We present cyclic and non-cyclic surface code simulations exhibiting this discrepancy, and analytic formulae precisely reproducing the observed behavior in the limit of low physical error. These asymptotic formulae are then used to prove that the underestimate is exponentially large in the code distance $d$ at any fixed physical error rate $p$ below the threshold error rate $p_{\rm th}$.
\end{abstract}

\maketitle

Topological quantum error correction (TQEC) is the lowest overhead technique for achieving reliable large-scale quantum computation given a 2-D lattice of qubits with nearest neighbor interactions \cite{Fowl12h}. As such, accurate simulations of the performance of such codes under physically realistic assumptions are of great practical relevance. Much of the existing literature on TQEC uses cyclic boundaries \cite{Barr10,Brav11,Brav12,Andr12}. While this is sufficient to obtain the threshold error rate $p_{\rm th}$ of a given TQEC scheme, we show that this is not sufficient to obtain an accurate logical error rate at a given code distance $d$ and physical error rate $p<p_{\rm th}$. Indeed, we show that cyclic boundaries lead to an exponentially growing underestimate of the logical error rate with $d$.

Without loss of generality, we shall focus on the surface code \cite{Brav98,Denn02,Raus07,Raus07d,Fowl08} making use of minimum weight perfect matching for decoding \cite{Edmo65a,Edmo65b,Fowl11b,Fowl12c}. Our results are relevant to any periodic, fault-tolerant implementation of any planar TQEC code. An up-to-date and detailed review of the surface code can be found in \cite{Fowl12f}.

While the field of TQEC is complex, our presentation shall assume only basic quantum mechanics and otherwise be self-contained. We shall start with an introduction to the surface code.

\begin{figure}
\begin{center}
\includegraphics[width=80mm]{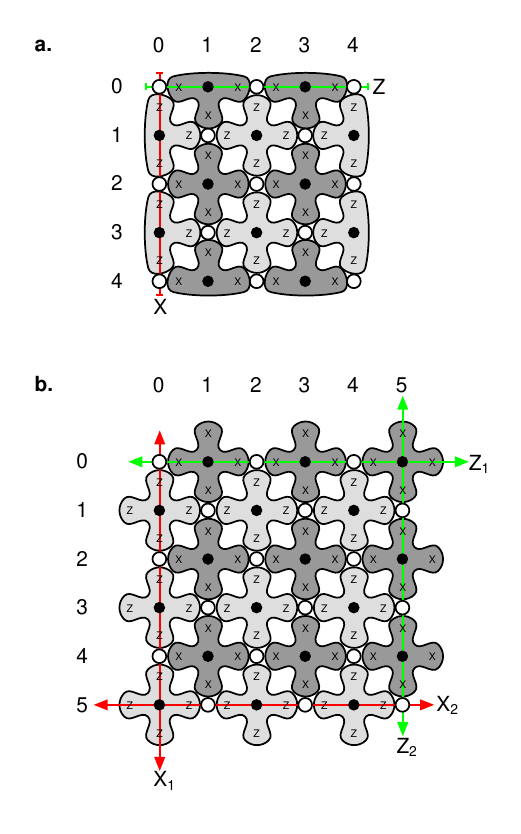}
\end{center}
\caption{Patterns of stabilizers of a distance $d=3$ surface code with {\bf a.}~non-cyclic boundaries, {\bf b.}~cyclic boundaries. Examples of logical operators of each type have been given. Each white and black dot represents a qubit. Light-colored bubbles around groups of $Z$ operators and dark-colored bubbles around groups of $X$ operators denote tensor products, with the extremities of each bubble touching the qubits operated on by each component of the tensor product.}\label{sc}
\end{figure}

The patterns of stabilizers \cite{Gott97} associated with distance $d=3$ non-cyclic and cyclic surface codes are shown in Fig.~\ref{sc}. A stabilizer of a state $\ket{\Psi}$ is simply an operator $A$ such that $A\ket{\Psi}=\ket{\Psi}$. We denote the Pauli matrices by $X$, $Y$ and $Z$. Each white and black dot in Fig.~\ref{sc} is a quantum bit (qubit), namely a two-level quantum system. The two states are denoted $\ket{0}=(1,0)^T$ and $\ket{1}=(0,1)^T$. Light-colored bubbles around groups of $Z$ operators and dark-colored bubbles around groups of $X$ operators denote tensor products, with the extremities of each bubble touching the qubits operated on by each component of the tensor product. Note that all stabilizers commute, so a simultaneous eigenstate of all stabilizers exists.

\begin{figure}
\begin{center}
\includegraphics[width=50mm]{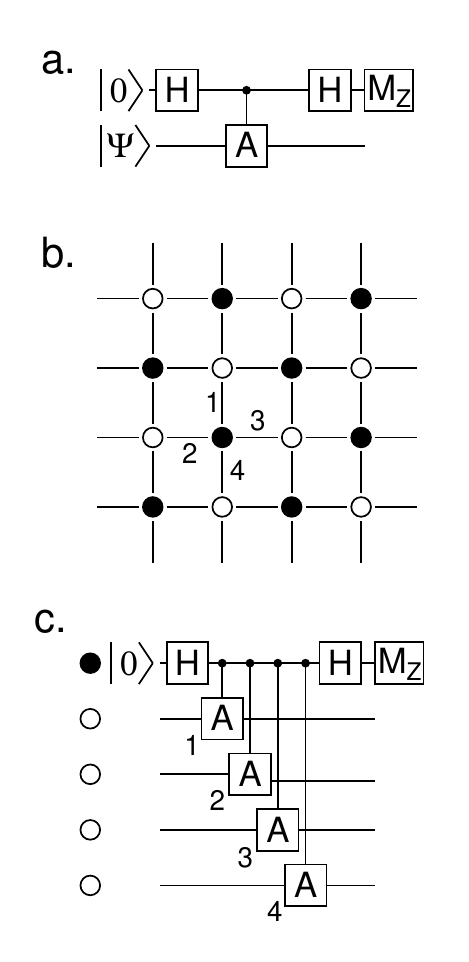}
\end{center}
\caption{{\bf a.}~A generic quantum circuit measuring the eigenvalue of any operator $A$ such that $A^2=I$. The $H$ (Hadamard) gate maps $\ket{0}\rightarrow (\ket{0}+\ket{1})/\sqrt{2}$ and $\ket{1}\rightarrow (\ket{0}-\ket{1})/\sqrt{2}$. The $M_Z$ gate represents projective measurement in the $Z$ basis. The central structure is a controlled-$A$ gate, meaning $A$ is applied only if the control qubit (the one touched by the dot) is $\ket{1}$. Time runs from left to right. {\bf b.}~Surface code qubit layout with numbered interactions. {\bf c.}~Generic quantum circuit measuring the stabilizers of the surface code ($A=X$ or $Z$). Numbers correspond to part b, and indicate a North, West, East, South interaction sequence. This sequence enables all stabilizers to be measured simultaneously.}\label{sequence}
\end{figure}

A generic quantum circuit measuring the eigenvalue of any operator $A$ such that $A^2=I$ is shown in Fig.~\ref{sequence}a. The $H$ (Hadamard) gate maps $\ket{0}\rightarrow (\ket{0}+\ket{1})/\sqrt{2}$ and $\ket{1}\rightarrow (\ket{0}-\ket{1})/\sqrt{2}$. The $M_Z$ gate represents projective measurement in the $Z$ basis. The central structure is a controlled-$A$ gate, meaning $A$ is applied only if the control qubit (the one touched by the dot) is $\ket{1}$. Time runs from left to right. Explicitly, the initial state is $\ket{0}\ket{\Psi}$. After the first Hadamard we will have $(\ket{0}+\ket{1})\ket{\Psi}/\sqrt{2}$. After controlled-$A$ we obtain $(\ket{0}\ket{\Psi}+\ket{1}A\ket{\Psi})/\sqrt{2}$. The second Hadamard gives $\ket{0}(\ket{\Psi}+A\ket{\Psi})/2+\ket{1}(\ket{\Psi}-A\ket{\Psi})/2$. If projective $Z$ basis measurement reports $\ket{0}$, we will have the +1 eigenstate of $A$, namely $(\ket{\Psi}+A\ket{\Psi})/\sqrt{2}$. Similarly, a report of $\ket{1}$ indicates the -1 eigenstate. Figs.~\ref{sequence}b--c show how to specialize this generic circuit to measure surface code stabilizers in a manner permitting simultaneous measurement of all stabilizers.

\begin{figure}
\begin{center}
\includegraphics[width=60mm]{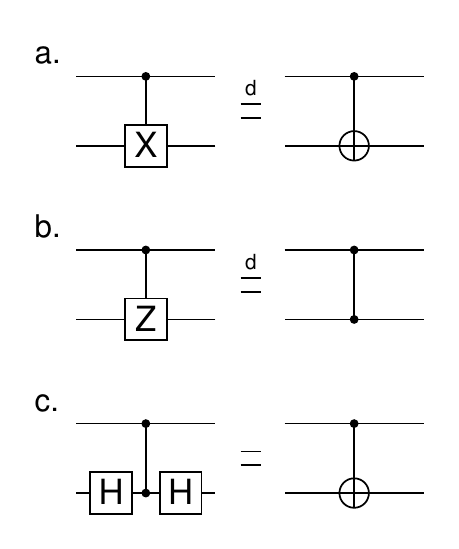}
\end{center}
\caption{{\bf a.}~Circuit symbol for controlled-$X$. {\bf b.}~Circuit symbol for controlled-$Z$. {\bf c.}~Circuit identity relating controlled-$X$ and controlled-$Z$.}\label{definitions}
\end{figure}

\begin{figure}
\begin{center}
\includegraphics[width=80mm]{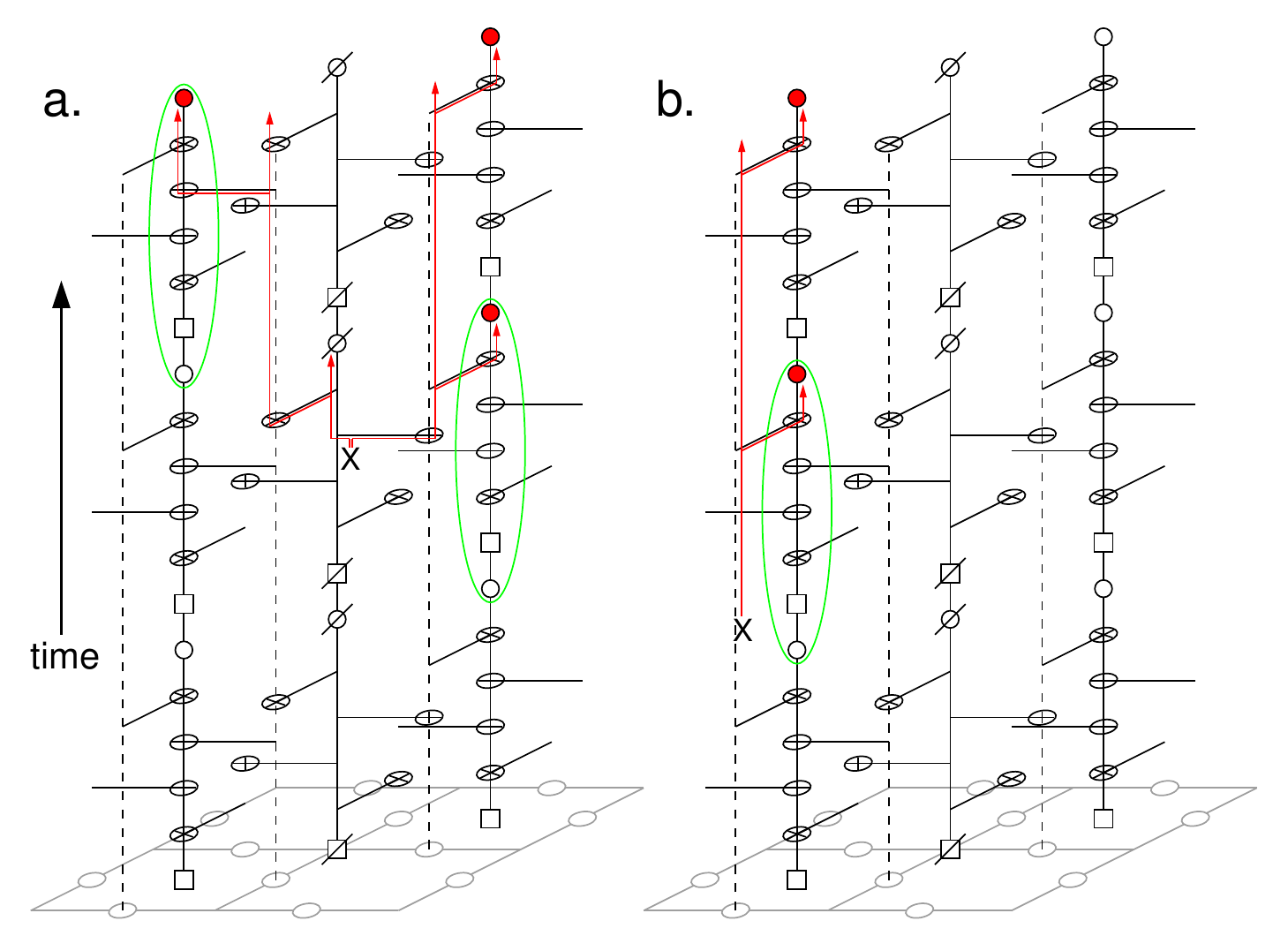}
\end{center}
\caption{2-D surface code (grey). Time runs vertically. Squares represent initialization to $\ket{0}$, circles represent $Z$ basis measurement. Slashed squares represent initialization to $\ket{+}$, slashed circles represent $X$ basis measurement (both achieved using Hadamard gates). {\bf a.}~A single error leading to a pair of detection events (green ellipses). A detection event is a sequential pair of measurements with differing value. Red lines show the paths of error propagation. {\bf b.}~An error leading to a single detection event due to proximity to a boundary of the lattice.}\label{syn_err}
\end{figure}

Using the circuit definitions and identity in Fig.~\ref{definitions}, a portion of the simultaneous stabilizer measurement procedure is shown in Fig.~\ref{syn_err}. This procedure has been designed to make use of only controlled-$X$ ($C_X$) interaction gates. Assuming control-target ordering, note that $C_X\ket{00}=\ket{00}$, whereas $C_X\ket{10}=\ket{11}$. This gives intuitive justification for the statement $C_XX_c\ket{\Psi}=X_cX_tC_X\ket{\Psi}$, namely $C_X$ copies $X$ errors from the control qubit to the target qubit, which can be verified via matrix multiplication.

A given stabilizer measurement, performed repeatedly in the absence of errors, will always return the same result, either +1 or -1. Two sequential measurements reporting different values are defined to be a detection event and indicate that at least one error has occurred nearby. We call the space-time coordinate of the beginning of the second measurement the space-time coordinate of the detection event. It is conceptually straightforward to analyze the propagation of all possible single errors and determine the total probability of any given pair of detection events due to all distinct single errors. We can visualize each such probability as a cylinder with space-time endpoints corresponding to the detection events and diameter proportional to the total probability. We call such a visualization a nest, and each cylinder a stick.

\begin{figure}
\begin{center}
\includegraphics[width=40mm]{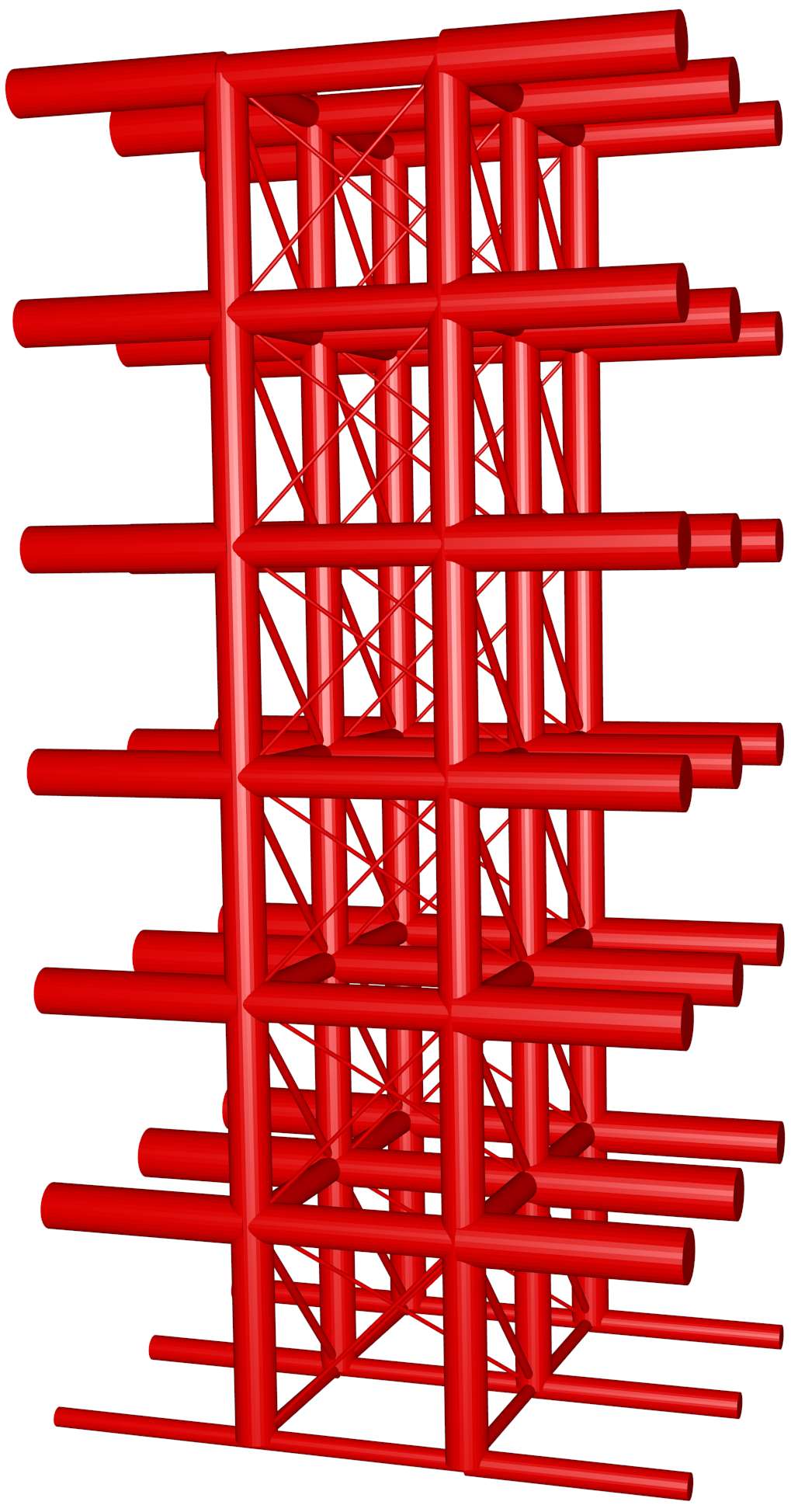}
\end{center}
\caption{Nest of 7 rounds of Fig.~\ref{sc}a fault-tolerant $X$ stabilizer measurements. Note that there are six columns of cylinders (sticks), with each column corresponding to a distinct $X$ stabilizer in Fig.~\ref{sc}a. Stabilizers are measured using the quantum circuits shown in Fig.~\ref{syn_err}. Under the assumption that the surface code is initially in the +1 eigenstate of all stabilizers, a physical measurement error during measurement of the bottom left $X$ stabilizer will lead to a -1 eigenstate being reported, generating a detection event. Under the assumption that this is the only error, the next measurement of this $X$ stabilizer will report +1, generating a second detection event associated with the physical measurement error. This pair of detection events along with the probability of the physical measurement error contributes to the diameter of the front, bottom, left vertical stick connecting the first two layers of the nest. By tracing the propagation of all possible errors in all seven rounds of the error detection circuitry, the complete nest is generated. This is done using our tool Autotune \cite{Fowl12d}.}\label{non-cyclic}
\end{figure}

\begin{figure}
\begin{center}
\includegraphics[width=40mm]{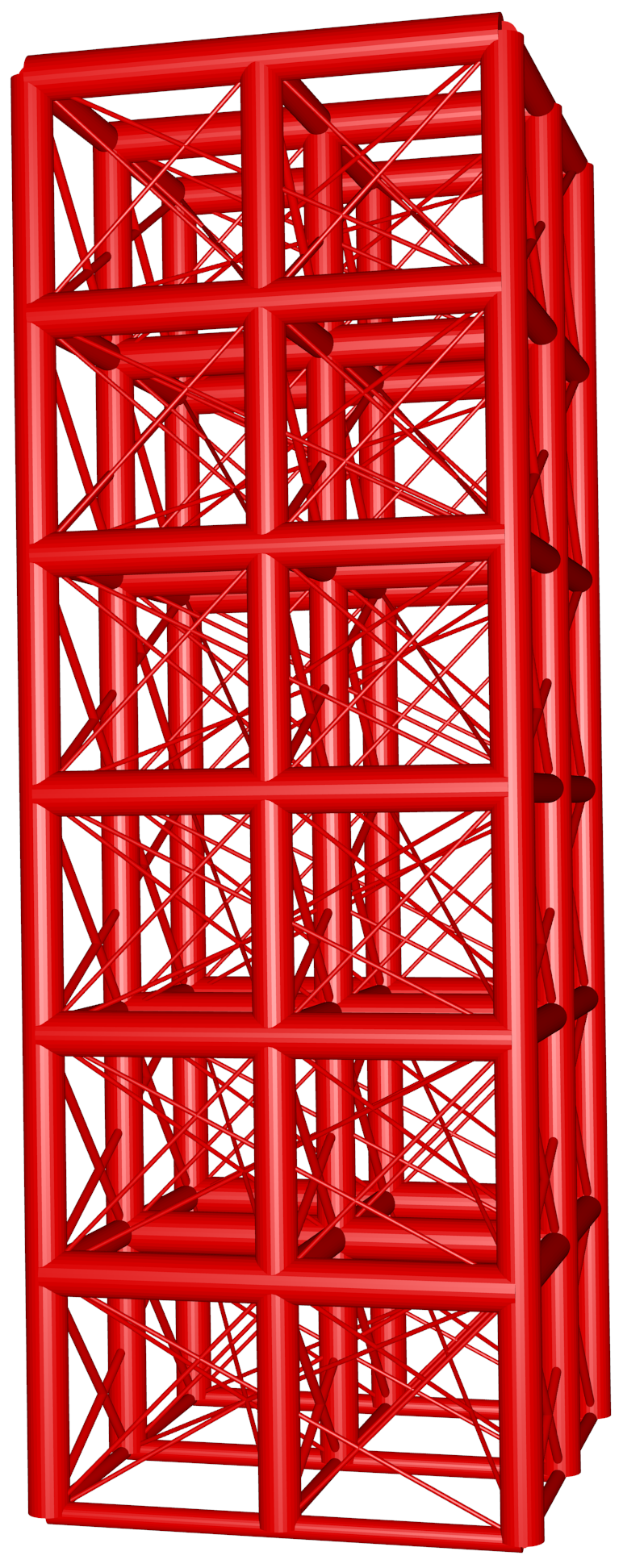}
\end{center}
\caption{Nest of 7 rounds of Fig.~\ref{sc}b fault-tolerant $X$ stabilizer measurements. Note that there are nine columns of cylinders (sticks), with each column corresponding to a distinct $X$ stabilizer in Fig.~\ref{sc}b. Stabilizers are measured using the quantum circuits shown in Fig.~\ref{syn_err}. In this case, not all sticks are visible as sticks connecting opposite extremities of the figure (cyclic boundaries) are indistinguishable from sticks connecting neighboring stabilizers in the interior. This means a detailed understanding of the geometry of the figure can only being gained by wading through the raw data contained in the Supplemental Material.}\label{cyclic}
\end{figure}

Nests for $X$ stabilizer measurement in the surface code assuming non-cyclic and cyclic boundaries are shown in Figs.~\ref{non-cyclic} and \ref{cyclic}, respectively. These figures were generated using our tool Autotune \cite{Fowl12d}. Due to the complexity of these figures and the impossibility of truly appreciating their structure from a single 2-D image, their raw data and Blender 3-D files have been included in the Supplemental Material. The reader seeking a deep understanding of this work needs to take the time to become familiar with the structure of these figures. Additional discussion of specific surface code error propagation details can be found in \cite{Wang11}, however this is not required reading.

The distance $d$ of a surface code is the length of the shortest chain of operators that commutes with all stabilizers and is not a product of stabilizers itself. In our simulations, we use a standard depolarizing error model, namely initialization prepares the wrong state with probability $p$, measurement reports the wrong value with probability $p$, single-qubit gates randomly apply one of $X$, $Y$ and $Z$ with total probability $p$, and two-qubit gates randomly apply one of the 15 non-trivial tensor products of $I$, $X$, $Y$ and $Z$ with total probability $p$. A logical error is a pattern of physical errors that is uncorrectable. A code with distance $d$ has logical states (e.g.~$\ket{0_L}$ and $\ket{1_L}$) that can only be interconverted by applying at least $d$ operators. This means, however, that the application of $\lceil (d+1)/2 \rceil$ operators can make you closer to a different logical state than the original, meaning correction will fail as correction takes you to the closest logical state. When $d$ is even, application of $d/2$ operators can leave you in an ambiguous state that can also cause correction to fail to return the quantum computer to the original state.

\begin{figure}
\begin{center}
\resizebox{85mm}{!}{\includegraphics[viewport=60 60 545 430, clip=true]{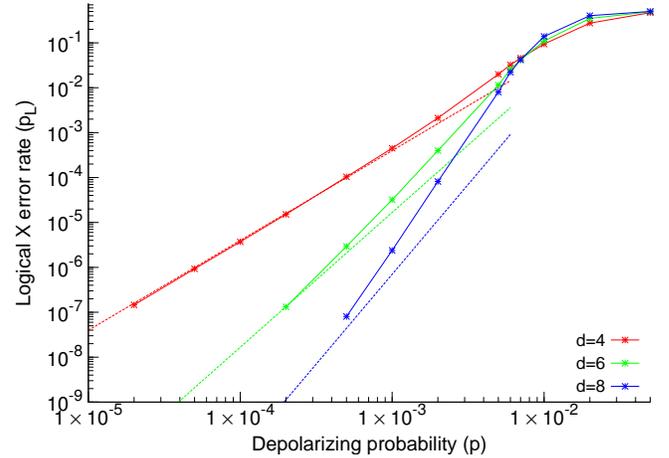}}
\end{center}
\caption{Probability of logical $X$ error as defined in Fig.~\ref{sc}a as a function of the depolarizing error rate $p$ for various distances $d$. The analytic forms of the dashed curves can be found in Table~\ref{comp}.}\label{logx}
\end{figure}

\begin{figure}
\begin{center}
\resizebox{85mm}{!}{\includegraphics[viewport=60 60 545 430, clip=true]{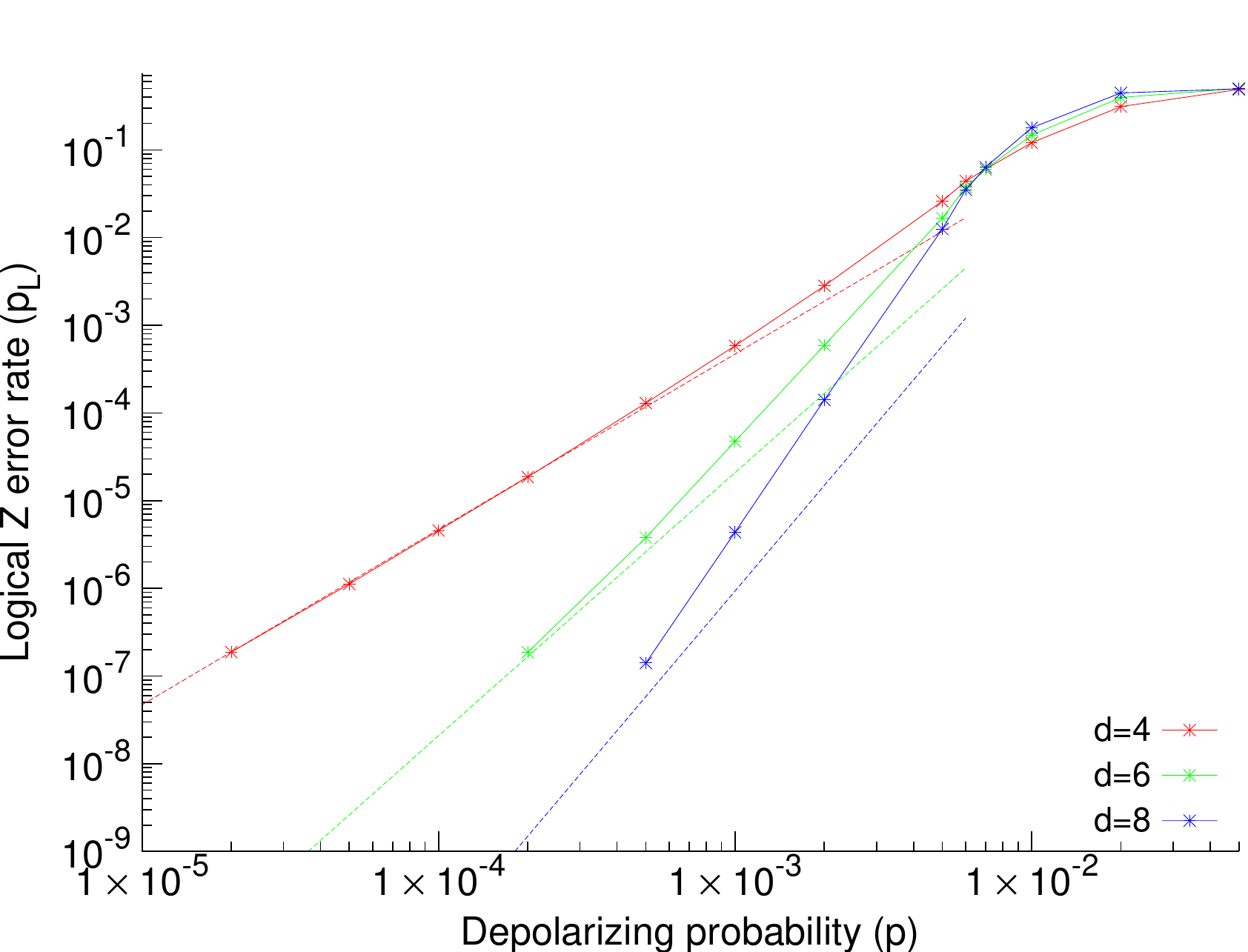}}
\end{center}
\caption{Probability of logical $Z$ error as defined in Fig.~\ref{sc}a as a function of the depolarizing error rate $p$ for various distances $d$. The analytic forms of the dashed curves can be found in Table~\ref{comp}.}\label{logz}
\end{figure}

\begin{figure}
\begin{center}
\resizebox{85mm}{!}{\includegraphics[viewport=60 60 545 430, clip=true]{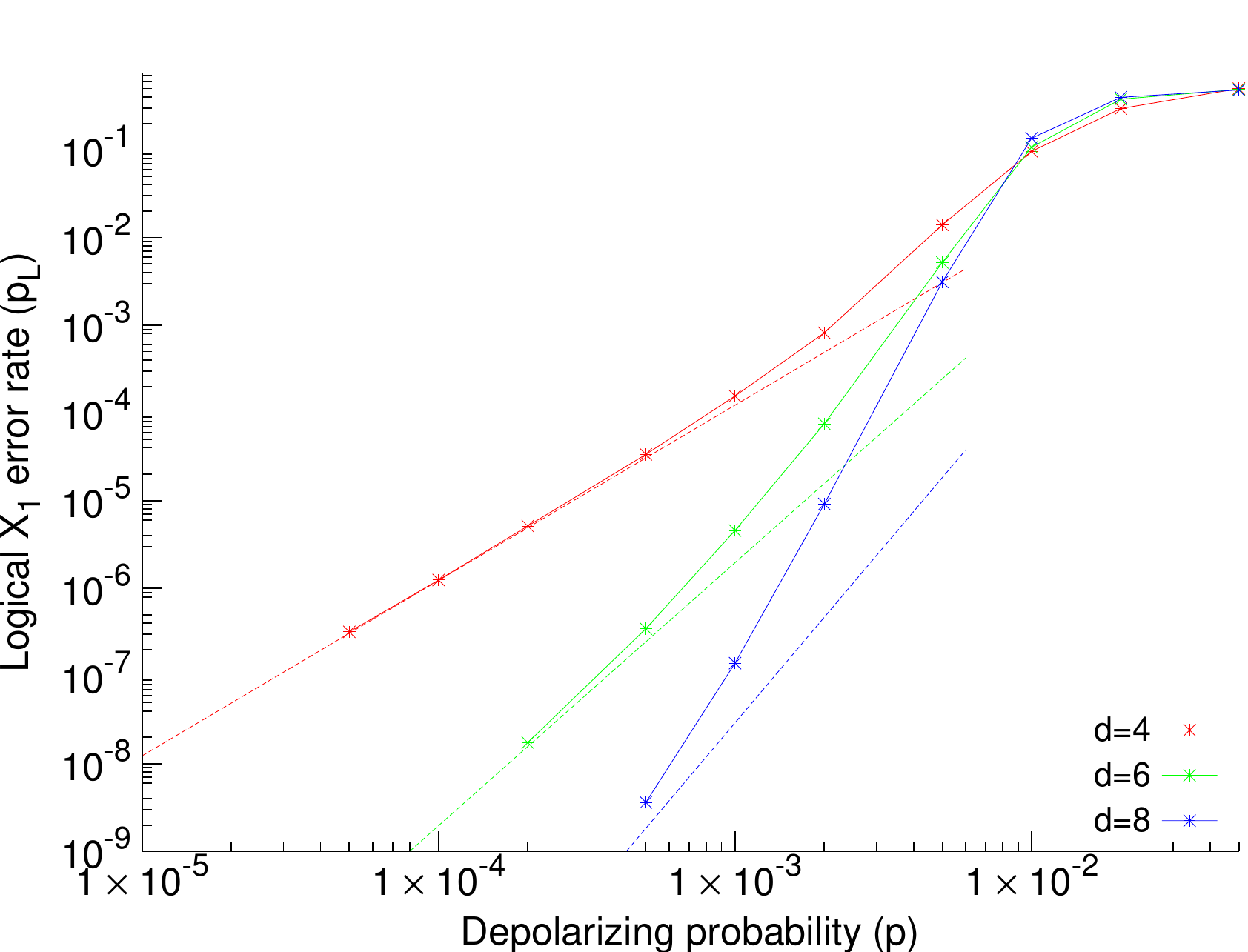}}
\end{center}
\caption{Probability of logical $X_1$ error as defined in Fig.~\ref{sc}b as a function of the depolarizing error rate $p$ for various distances $d$. The analytic forms of the dashed curves can be found in Table~\ref{comp}.}\label{logx1}
\end{figure}

\begin{figure}
\begin{center}
\resizebox{85mm}{!}{\includegraphics[viewport=60 60 545 430, clip=true]{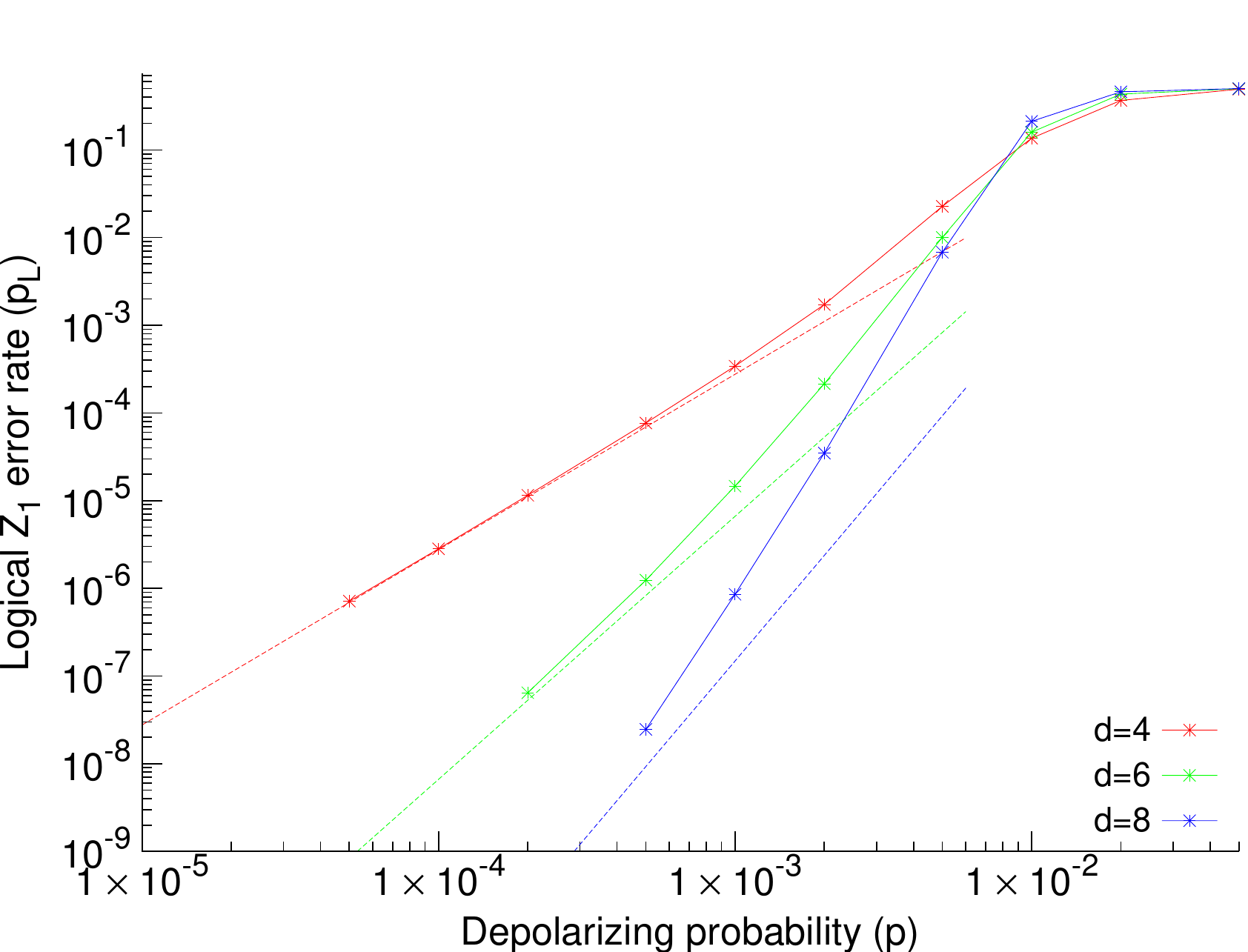}}
\end{center}
\caption{Probability of logical $Z_1$ error as defined in Fig.~\ref{sc}b as a function of the depolarizing error rate $p$ for various distances $d$. The analytic forms of the dashed curves can be found in Table~\ref{comp}.}\label{logz1}
\end{figure}

\begin{figure}
\begin{center}
\resizebox{85mm}{!}{\includegraphics[viewport=60 60 545 430, clip=true]{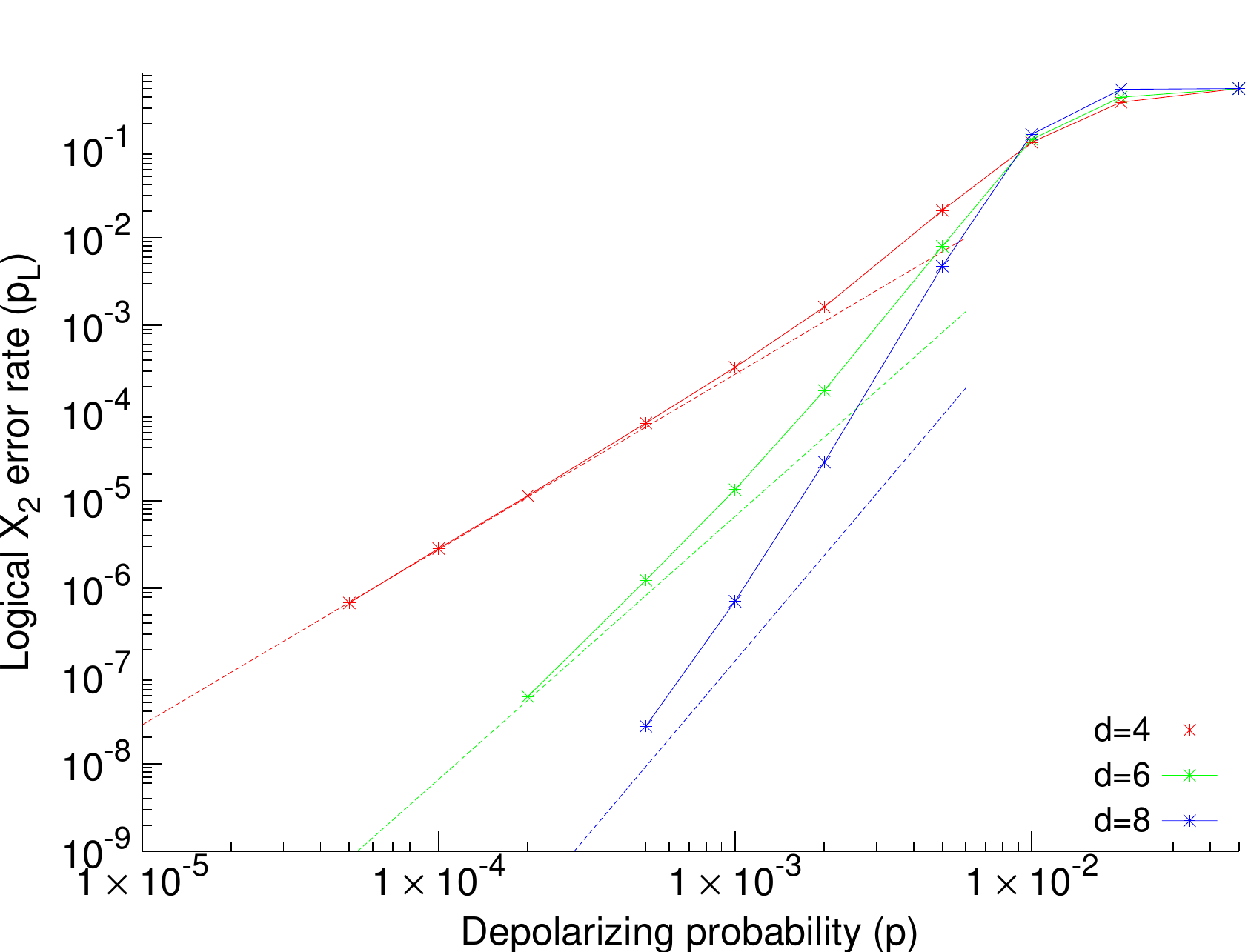}}
\end{center}
\caption{Probability of logical $X_2$ error as defined in Fig.~\ref{sc}b as a function of the depolarizing error rate $p$ for various distances $d$. The analytic forms of the dashed curves are identical to those of Fig.~\ref{logz1} and can be found in Table~\ref{comp}.}\label{logx2}
\end{figure}

\begin{figure}
\begin{center}
\resizebox{85mm}{!}{\includegraphics[viewport=60 60 545 430, clip=true]{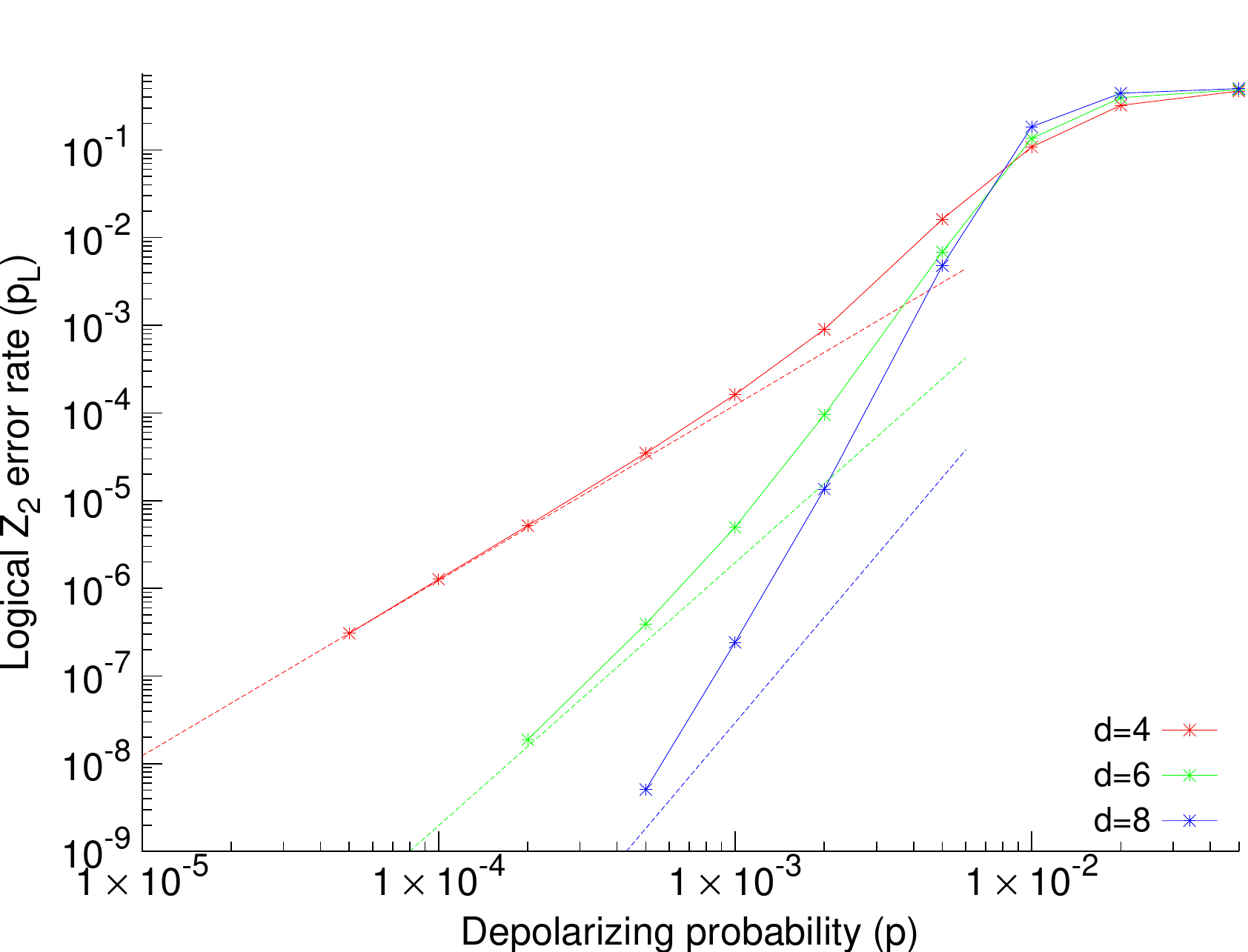}}
\end{center}
\caption{Probability of logical $Z_2$ error as defined in Fig.~\ref{sc}b as a function of the depolarizing error rate $p$ for various distances $d$. The analytic forms of the dashed curves are identical to those of Fig.~\ref{logx1} and can be found in Table~\ref{comp}.}\label{logz2}
\end{figure}

The probability of each type of logical error per round of error detection for various distances $d$ and non-cyclic and cyclic boundaries as a function of the physical depolarizing error rate $p$ is shown in Figs.~\ref{logx}--\ref{logz2}. It can be seen that the logical error rates of the cyclic case are significantly lower, particularly logical $X_1$ and $Z_2$ errors. We can determine the low $p$ asymptotic forms of the curves in Figs.~\ref{logx}--\ref{logz} using the work of \cite{Fowl12g}. These analytic expressions (simple equations of the form $Ap^{d/2}$) are shown as dashed lines and detailed in Table~\ref{comp} and were derived completely independently of the simulations. The agreement with the simulation data is perfect to numerical precision, implying the processes leading to failure in the non-cyclic surface code are well understood. For the purposes of this paper, it is only important to note that the asymptotic form is as it should be, namely at least $d/2$ errors, each of which has probability $O(p)$, are required to cause failure.

\begin{table}
\begin{tabular}{c|c|c|c|c}
$d$ & $A_{X}$ & $A_{Z}$ & $A_{X_1}$ & $A_{Z_1}$ \\
\hline
4 & $3.97\times 10^2$ & $4.70\times 10^2$ & $1.23\times 10^2$ & $2.76\times 10^2$ \\
6 & $1.67\times 10^4$ & $2.09\times 10^4$ & $1.97\times 10^3$ & $6.64\times 10^3$ \\
8 & $7.02\times 10^5$ & $9.34\times 10^5$ & $2.94\times 10^4$ & $1.49\times 10^5$ \\
10 & $2.93\times 10^7$ & $4.18\times 10^7$ & $4.23\times 10^5$ & $3.21\times 10^6$ \\
\end{tabular}
\caption{Low $p$ analytic asymptotic formulae derived from first principles, not simulation. Each entry represents a curve of the form $Ap^{d/2}$. The definitions of the logical operators $X$, $Z$, $X_1$, $Z_1$ can be found in Fig.~\ref{sc}.}
\label{comp}
\end{table}

As the non-cyclic code distance $d$ increases, the number of minimum stick topologically nontrivial paths connecting opposing boundaries grows exponentially (consider larger versions of Fig.~\ref{non-cyclic}). Such paths can be associated with logical operators of minimum weight $d$, each of which crudely speaking increases the probability of logical error. More precise discussion can be found in \cite{Fowl12g}. By contrast, in a distance $d$ surface code with cyclic boundaries, there are precisely $d$ minimum stick topologically nontrivial cyclic paths. Diagonal sticks never zigzag and hence never create additional cyclic paths.

While the most concrete way to verify that diagonal sticks never zigzag is to pore through the raw data in the Supplemental Material, the reason for this behavior can be understood intuitively. Consider any cyclic boundary, transversely invariant, periodic, fault-tolerant TQEC with the property that any single error generates two detection events. Choose any data qubit. By the assumption that any single error generates two detection events, any such data qubit will be a member of precisely two stabilizers of each type ($X$ and $Z$). By the assumption of fault-tolerance, the data qubit cannot simultaneously interact with syndrome qubits associated with different stabilizers. Without loss of generality, let us assume that we have a $Z$ stabilizer to the left and right of the data qubit. By the assumption of transversal invariance, all data qubits with a $Z$ stabilizer to the left and right must all interact with one or the other first. Without loss of generality, let us assume that the data qubit interacts with the syndrome qubit to its right first. If an $X$ error occurs after interaction with the right syndrome qubit, this error will be detected by the left syndrome qubit in this round of error detection, but not by the right syndrome qubit until the next round of error detection. This will result in a diagonal stick with endpoints separated by one unit of space and one unit of time. No matter when an $X$ error occurs on this data qubit, the right end of the stick will never be lower than the left. In this example, only horizontal or upward diagonal to the right sticks will be generated. By systematically considering all gates and qubits, similar arguments can always be constructed. In short, diagonal sticks never zigzag.

Logical errors will occur 50\% of the time when physical errors occur corresponding to half the sticks in a given minimum weight logical operator. This implies the low $p$ asymptotic logical error rate, for even distances, will be
\begin{equation}
p_L=\frac{1}{2}d\left( \begin{array}{c} d \\ d/2 \end{array} \right)\epsilon^{d/2},
\label{cyclic_eq}
\end{equation}
where $\epsilon$ is the probability of a stick within a minimum stick logical operator. Note that the symmetry of a cyclic boundary surface code implies that all such sticks will have the same probability.

From the Supplemental Material, we find that a $Z_1$ logical operator (left to right) is associated with sticks of probability $\epsilon=4.8p$, for example by considering the $(4,1,6)$ to $(4,3,6)$ stick (first two numbers correspond to the coordinates in Fig.~\ref{sc}b, entry is marked with a * in the Supplemental Material, $\epsilon$ listed there is for $p=0.04$). A $Z_2$ logical operator (top to bottom) is associated with sticks of probability $\epsilon=3.2p$. Inserting these expressions into Eq.~\ref{cyclic_eq} and noting that by symmetry the logical $X_1$ and $Z_2$ error rates will be the same, as will the logical $X_2$ and $Z_1$ error rates, we can see in Table~\ref{comp} that at distance $d=10$ the cyclic logical $Z$ error rate is already over a factor of 10 too low and the cyclic logical $X$ error rate nearly 2 orders of magnitude too low. The degree of underestimation grows exponentially with $d$.

It might be thought that the asymptotic expressions of Table~\ref{comp} are only valid approximations for very low $p$ and that the approximation grows worse as the code distance $d$ increases. This is not the case. In any given round of non-cyclic error detection, we can choose $d$ boundary sticks to associate endpoints of logical operators with. A logical operator must be a non-self-intersecting paths of sticks connecting to anywhere on the opposing boundary. A logical operator must contain at least $d$ sticks to connect opposing boundaries, but can contain arbitrarily many more. Given no point in the nests of the surface code, cyclic or otherwise, is associated with more than 12 sticks, the number of logical operators associated with a given round of error detection is upper bounded by $\sum_{m=d}^\infty d 11^{m-2}$. A logical error will be associated with a logical operator if at least $\lceil m/2 \rceil$ of its sticks are associated with physical errors.

Given a particular path of length $m$, the probability of at least $\lceil m/2 \rceil$ of its lines being associated with errors is
\begin{eqnarray}
\sum_{i=\lceil \frac{m}{2} \rceil}^m \left( \begin{array}{c} m \\ i \end{array} \right)\epsilon^i & \leq & \sum_{i=\lceil \frac{m}{2} \rceil}^m \left( \begin{array}{c} m \\ \lceil \frac{m}{2} \rceil \end{array} \right)\epsilon^i \\
& = & \left( \begin{array}{c} m \\ \lceil \frac{m}{2} \rceil \end{array} \right)\epsilon^{\lceil \frac{m}{2} \rceil}\sum_{i=0}^{m-\lceil \frac{m}{2} \rceil}\epsilon^i \\
& \leq & \left( \begin{array}{c} m \\ \lceil \frac{m}{2} \rceil \end{array} \right)\epsilon^{\lceil \frac{m}{2} \rceil}\frac{1}{1-\epsilon} \\
& \leq & \epsilon^{\lceil \frac{m}{2} \rceil}\sum_{i=0}^m \left( \begin{array}{c} m \\ i \end{array} \right) \\
& = & 2^m \epsilon^{\lceil \frac{m}{2} \rceil}
\end{eqnarray}
The total probability of logical error per round of error detection is therefore upper bounded by
\begin{eqnarray}
& & \sum_{m=d}^\infty d 11^{m-2} 2^m \epsilon^{\lceil \frac{m}{2} \rceil} \\
& = & d 11^{d-2} 2^d \sum_{m=d}^\infty 11^{m-d} 2^{m-d} \epsilon^{\lceil \frac{m}{2} \rceil}
\end{eqnarray}
Restricting to even $d$ (for convenience) gives
\begin{eqnarray}
& & d \frac{22^d}{121}\epsilon^{d/2} \sum_{m=0}^\infty 22^m \epsilon^{\lceil \frac{m}{2} \rceil} \\
& = & d \frac{22^d}{121}\epsilon^{d/2} \sum_{m=0}^\infty (22^{2m} + 22^{2m+1}) \epsilon^m \\
& < & 44d \frac{22^d}{121}\epsilon^{d/2} \sum_{m=0}^\infty 484^m \epsilon^m
\end{eqnarray}
Writing the prefactor as $B$ yields
\begin{eqnarray}
& & B\sum_{m=0}^\infty 484^m \epsilon^m \\
& = & B(1 + \sum_{m=1}^\infty 484^m \epsilon^m) \\
& = & B(1 + \frac{484\epsilon}{1-484\epsilon})
\end{eqnarray}

The leading order $\epsilon^{d/2}$ behavior therefore provably dominates for $\epsilon\ll 1/484$, for arbitrary $d$. Since our argument was not particularly tight, leading order behavior actually dominates at significantly higher values. This implies that there is a constant and reasonably high value of $p$ at which leading order error processes dominate for arbitrarily large $d$. This makes sense, as at some constant low density of errors, the overwhelmingly most likely type of logical error at any arbitrarily large $d$ will be one consisting of the minimum possible number of errors.

Furthermore, given we can choose some value of $p$ at which our asymptotics apply with high accuracy for arbitrary $d$, the exponential separation of the cyclic and non-cyclic logical error rates holds at any value $p$ all the way up to the threshold error rate $p_{\rm th}$ as the logical error rate curves must pivot at the threshold and stretch down to meet the asymptotic curves.

In summary, we have shown that one of the most commonly used simplifying assumptions in physics and mathematics, namely cyclic boundaries, cannot be used if one wishes to accurately study the logical error rates of planar topological codes. The level of inaccuracy introduced by this seemingly harmless assumption is exponentially large in the code distance $d$. This implies that cyclic boundaries should be avoided when performing accurate studies of practical topological quantum error correction. We hope that this work encourages the quantum information community to seriously question the accuracy of results derived using unphysical assumptions. Two common assumptions that we feel deserve particular scrutiny are assuming perfect multi-body quantum measurements, and assuming arbitrarily long range coherent interactions without time, overhead, or error rate penalty.

\section{Acknowledgements}

This research was conducted by the Australian Research Council Centre of Excellence for Quantum Computation and Communication Technology (project number CE110001027). Supported by the Intelligence Advanced Research Projects Activity (IARPA) via Department of Interior National Business Center contract number D11PC20166.  The U.S. Government is authorized to reproduce and distribute reprints for Governmental purposes notwithstanding any copyright annotation thereon.  Disclaimer: The views and conclusions contained herein are those of the authors and should not be interpreted as necessarily representing the official policies or endorsements, either expressed or implied, of IARPA, DoI/NBC, or the U.S. Government.

\bibliography{../References}

\clearpage

\section{Supplemental Material}

We include below the raw data for the nests of Fig.~5 and Fig.~6. The initial two groups of three numbers in each line are the space-time coordinates of a cylinder, with the $i$, $j$ coordinates corresponding to the numbers in Fig.~1, and the $t$ coordinate corresponding to the number of rounds of error detection since the first, doubled. The time $t$ therefore takes the values $\{0, 2, \ldots, 12\}$. The doubling of the time coordinate is purely to improve the symmetry of the nests. The final number in each line is the diameter of each cylinder and is the total probability of detection events at the end points of each cylinder from single errors when using a standard balanced depolarizing error model of strength $p=0.04$. This value of $p$ was chosen simply to produce a clear nest. The diameter of each cylinder scales linearly with $p$.

\subsection{Data for Fig.~5 (non-cyclic)}

\begin{verbatim}
[
 [(4, 3, 12), (4, 3, 10), 0.1920499],
 [(4, 3, 12), (4, 1, 10), 0.0213650],
 [(4, 3, 12), (4, 5, 12), 0.2402681],
 [(4, 3, 12), (2, 3, 10), 0.0426811],
 [(4, 1, 12), (4, 3, 12), 0.1867471],
 [(4, 1, 12), (4, 1, 10), 0.1920499],
 [(4, 1, 12), (4, -1, 12), 0.2135253],
 [(4, 1, 12), (2, 3, 10), 0.0213650],
 [(4, 1, 12), (2, 1, 10), 0.0426811],
 [(2, 3, 12), (4, 3, 12), 0.1282210],
 [(2, 3, 12), (2, 3, 10), 0.1762239],
 [(2, 3, 12), (2, 1, 10), 0.0213650],
 [(2, 3, 12), (2, 5, 12), 0.2456125],
 [(2, 3, 12), (0, 3, 10), 0.0426811],
 [(2, 1, 12), (4, 1, 12), 0.1282210],
 [(2, 1, 12), (2, 3, 12), 0.1920499],
 [(2, 1, 12), (2, 1, 10), 0.1762239],
 [(2, 1, 12), (2, -1, 12), 0.2456125],
 [(2, 1, 12), (0, 3, 10), 0.0213650],
 [(2, 1, 12), (0, 1, 10), 0.0426811],
 [(0, 3, 12), (2, 3, 12), 0.1282210],
 [(0, 3, 12), (0, 3, 10), 0.1920499],
 [(0, 3, 12), (0, 1, 10), 0.0213650],
 [(0, 3, 12), (0, 5, 12), 0.2135253],
 [(0, 1, 12), (2, 1, 12), 0.1282210],
 [(0, 1, 12), (0, 3, 12), 0.1867471],
 [(0, 1, 12), (0, 1, 10), 0.1920499],
 [(0, 1, 12), (0, -1, 12), 0.2402681],
 [(4, 3, 10), (4, 3, 8), 0.1920499],
 [(4, 3, 10), (4, 1, 8), 0.0213650],
 [(4, 3, 10), (4, 5, 10), 0.2402681],
 [(4, 3, 10), (2, 3, 8), 0.0426811],
 [(4, 1, 10), (4, 3, 10), 0.1867471],
 [(4, 1, 10), (4, 1, 8), 0.1920499],
 [(4, 1, 10), (4, -1, 10), 0.2135253],
 [(4, 1, 10), (2, 3, 8), 0.0213650],
 [(4, 1, 10), (2, 1, 8), 0.0426811],
 [(2, 3, 10), (4, 3, 10), 0.1282210],
 [(2, 3, 10), (2, 3, 8), 0.1762239],
 [(2, 3, 10), (2, 1, 8), 0.0213650],
 [(2, 3, 10), (2, 5, 10), 0.2456125],
 [(2, 3, 10), (0, 3, 8), 0.0426811],
 [(2, 1, 10), (4, 1, 10), 0.1282210],
 [(2, 1, 10), (2, 3, 10), 0.1920499],
 [(2, 1, 10), (2, 1, 8), 0.1762239],
 [(2, 1, 10), (2, -1, 10), 0.2456125],
 [(2, 1, 10), (0, 3, 8), 0.0213650],
 [(2, 1, 10), (0, 1, 8), 0.0426811],
 [(0, 3, 10), (2, 3, 10), 0.1282210],
 [(0, 3, 10), (0, 3, 8), 0.1920499],
 [(0, 3, 10), (0, 1, 8), 0.0213650],
 [(0, 3, 10), (0, 5, 10), 0.2135253],
 [(0, 1, 10), (2, 1, 10), 0.1282210],
 [(0, 1, 10), (0, 3, 10), 0.1867471],
 [(0, 1, 10), (0, 1, 8), 0.1920499],
 [(0, 1, 10), (0, -1, 10), 0.2402681],
 [(4, 3, 8), (4, 3, 6), 0.1920499],
 [(4, 3, 8), (4, 1, 6), 0.0213650],
 [(4, 3, 8), (4, 5, 8), 0.2402681],
 [(4, 3, 8), (2, 3, 6), 0.0426811],
 [(4, 1, 8), (4, 3, 8), 0.1867471],
 [(4, 1, 8), (4, 1, 6), 0.1920499],
 [(4, 1, 8), (4, -1, 8), 0.2135253],
 [(4, 1, 8), (2, 3, 6), 0.0213650],
 [(4, 1, 8), (2, 1, 6), 0.0426811],
 [(2, 3, 8), (4, 3, 8), 0.1282210],
 [(2, 3, 8), (2, 3, 6), 0.1762239],
 [(2, 3, 8), (2, 1, 6), 0.0213650],
 [(2, 3, 8), (2, 5, 8), 0.2456125],
 [(2, 3, 8), (0, 3, 6), 0.0426811],
 [(2, 1, 8), (4, 1, 8), 0.1282210],
 [(2, 1, 8), (2, 3, 8), 0.1920499],
 [(2, 1, 8), (2, 1, 6), 0.1762239],
 [(2, 1, 8), (2, -1, 8), 0.2456125],
 [(2, 1, 8), (0, 3, 6), 0.0213650],
 [(2, 1, 8), (0, 1, 6), 0.0426811],
 [(0, 3, 8), (2, 3, 8), 0.1282210],
 [(0, 3, 8), (0, 3, 6), 0.1920499],
 [(0, 3, 8), (0, 1, 6), 0.0213650],
 [(0, 3, 8), (0, 5, 8), 0.2135253],
 [(0, 1, 8), (2, 1, 8), 0.1282210],
 [(0, 1, 8), (0, 3, 8), 0.1867471],
 [(0, 1, 8), (0, 1, 6), 0.1920499],
 [(0, 1, 8), (0, -1, 8), 0.2402681],
 [(4, 3, 6), (4, 3, 4), 0.1920499],
 [(4, 3, 6), (4, 1, 4), 0.0213650],
 [(4, 3, 6), (4, 5, 6), 0.2402681],
 [(4, 3, 6), (2, 3, 4), 0.0426811],
 [(4, 1, 6), (4, 3, 6), 0.1867471],
 [(4, 1, 6), (4, 1, 4), 0.1920499],
 [(4, 1, 6), (4, -1, 6), 0.2135253],
 [(4, 1, 6), (2, 3, 4), 0.0213650],
 [(4, 1, 6), (2, 1, 4), 0.0426811],
 [(2, 3, 6), (4, 3, 6), 0.1282210],
 [(2, 3, 6), (2, 3, 4), 0.1762239],
 [(2, 3, 6), (2, 1, 4), 0.0213650],
 [(2, 3, 6), (2, 5, 6), 0.2456125],
 [(2, 3, 6), (0, 3, 4), 0.0426811],
 [(2, 1, 6), (4, 1, 6), 0.1282210],
 [(2, 1, 6), (2, 3, 6), 0.1920499],
 [(2, 1, 6), (2, 1, 4), 0.1762239],
 [(2, 1, 6), (2, -1, 6), 0.2456125],
 [(2, 1, 6), (0, 3, 4), 0.0213650],
 [(2, 1, 6), (0, 1, 4), 0.0426811],
 [(0, 3, 6), (2, 3, 6), 0.1282210],
 [(0, 3, 6), (0, 3, 4), 0.1920499],
 [(0, 3, 6), (0, 1, 4), 0.0213650],
 [(0, 3, 6), (0, 5, 6), 0.2135253],
 [(0, 1, 6), (2, 1, 6), 0.1282210],
 [(0, 1, 6), (0, 3, 6), 0.1867471],
 [(0, 1, 6), (0, 1, 4), 0.1920499],
 [(0, 1, 6), (0, -1, 6), 0.2402681],
 [(4, 3, 4), (4, 3, 2), 0.1920499],
 [(4, 3, 4), (4, 1, 2), 0.0213650],
 [(4, 3, 4), (4, 5, 4), 0.2402681],
 [(4, 3, 4), (2, 3, 2), 0.0426811],
 [(4, 1, 4), (4, 3, 4), 0.1867471],
 [(4, 1, 4), (4, 1, 2), 0.1920499],
 [(4, 1, 4), (4, -1, 4), 0.2135253],
 [(4, 1, 4), (2, 3, 2), 0.0213650],
 [(4, 1, 4), (2, 1, 2), 0.0426811],
 [(2, 3, 4), (4, 3, 4), 0.1282210],
 [(2, 3, 4), (2, 3, 2), 0.1762239],
 [(2, 3, 4), (2, 1, 2), 0.0213650],
 [(2, 3, 4), (2, 5, 4), 0.2456125],
 [(2, 3, 4), (0, 3, 2), 0.0426811],
 [(2, 1, 4), (4, 1, 4), 0.1282210],
 [(2, 1, 4), (2, 3, 4), 0.1920499],
 [(2, 1, 4), (2, 1, 2), 0.1762239],
 [(2, 1, 4), (2, -1, 4), 0.2456125],
 [(2, 1, 4), (0, 3, 2), 0.0213650],
 [(2, 1, 4), (0, 1, 2), 0.0426811],
 [(0, 3, 4), (2, 3, 4), 0.1282210],
 [(0, 3, 4), (0, 3, 2), 0.1920499],
 [(0, 3, 4), (0, 1, 2), 0.0213650],
 [(0, 3, 4), (0, 5, 4), 0.2135253],
 [(0, 1, 4), (2, 1, 4), 0.1282210],
 [(0, 1, 4), (0, 3, 4), 0.1867471],
 [(0, 1, 4), (0, 1, 2), 0.1920499],
 [(0, 1, 4), (0, -1, 4), 0.2402681],
 [(4, 3, 2), (4, 3, 0), 0.1920499],
 [(4, 3, 2), (4, 1, 0), 0.0213650],
 [(4, 3, 2), (4, 5, 2), 0.2402681],
 [(4, 3, 2), (2, 3, 0), 0.0426811],
 [(4, 1, 2), (4, 3, 2), 0.1867471],
 [(4, 1, 2), (4, 1, 0), 0.1920499],
 [(4, 1, 2), (4, -1, 2), 0.2135253],
 [(4, 1, 2), (2, 3, 0), 0.0213650],
 [(4, 1, 2), (2, 1, 0), 0.0426811],
 [(2, 3, 2), (4, 3, 2), 0.1282210],
 [(2, 3, 2), (2, 3, 0), 0.1762239],
 [(2, 3, 2), (2, 1, 0), 0.0213650],
 [(2, 3, 2), (2, 5, 2), 0.2456125],
 [(2, 3, 2), (0, 3, 0), 0.0426811],
 [(2, 1, 2), (4, 1, 2), 0.1282210],
 [(2, 1, 2), (2, 3, 2), 0.1920499],
 [(2, 1, 2), (2, 1, 0), 0.1762239],
 [(2, 1, 2), (2, -1, 2), 0.2456125],
 [(2, 1, 2), (0, 3, 0), 0.0213650],
 [(2, 1, 2), (0, 1, 0), 0.0426811],
 [(0, 3, 2), (2, 3, 2), 0.1282210],
 [(0, 3, 2), (0, 3, 0), 0.1920499],
 [(0, 3, 2), (0, 1, 0), 0.0213650],
 [(0, 3, 2), (0, 5, 2), 0.2135253],
 [(0, 1, 2), (2, 1, 2), 0.1282210],
 [(0, 1, 2), (0, 3, 2), 0.1867471],
 [(0, 1, 2), (0, 1, 0), 0.1920499],
 [(0, 1, 2), (0, -1, 2), 0.2402681],
 [(4, 3, 0), (4, 5, 0), 0.1174198],
 [(4, 1, 0), (4, 3, 0), 0.0907179],
 [(4, 1, 0), (4, -1, 0), 0.0907179],
 [(2, 3, 0), (4, 3, 0), 0.0640559],
 [(2, 3, 0), (2, 5, 0), 0.1227016],
 [(2, 1, 0), (4, 1, 0), 0.0640559],
 [(2, 1, 0), (2, 3, 0), 0.0961352],
 [(2, 1, 0), (2, -1, 0), 0.1227016],
 [(0, 3, 0), (2, 3, 0), 0.0640559],
 [(0, 3, 0), (0, 5, 0), 0.1227016],
 [(0, 1, 0), (2, 1, 0), 0.0640559],
 [(0, 1, 0), (0, 3, 0), 0.0961352],
 [(0, 1, 0), (0, -1, 0), 0.1227016]
]
\end{verbatim}

\subsection{Data for Fig.~6 (cyclic)}

\begin{verbatim}
[
 [(4, 5, 12), (4, 5, 10), 0.1760737],
 [(4, 5, 12), (4, 3, 10), 0.0213468],
 [(4, 5, 12), (2, 5, 10), 0.0427301],
 [(4, 5, 12), (2, 1, 10), 0.0213468],
 [(4, 3, 12), (4, 5, 12), 0.1922704],
 [(4, 3, 12), (4, 3, 10), 0.1760737],
 [(4, 3, 12), (4, 1, 10), 0.0213468],
 [(4, 3, 12), (2, 5, 10), 0.0213468],
 [(4, 3, 12), (2, 3, 10), 0.0427301],
 [(4, 1, 12), (4, 5, 12), 0.1922704],
 [(4, 1, 12), (4, 5, 10), 0.0213468],
 [(4, 1, 12), (4, 3, 12), 0.1922704],
 [(4, 1, 12), (4, 1, 10), 0.1760737],
 [(4, 1, 12), (2, 3, 10), 0.0213468],
 [(4, 1, 12), (2, 1, 10), 0.0427301],
 [(2, 5, 12), (4, 5, 12), 0.1281117],
 [(2, 5, 12), (2, 5, 10), 0.1760737],
 [(2, 5, 12), (2, 3, 10), 0.0213468],
 [(2, 5, 12), (0, 5, 10), 0.0427301],
 [(2, 5, 12), (0, 1, 10), 0.0213468],
 [(2, 3, 12), (4, 3, 12), 0.1281117],
 [(2, 3, 12), (2, 5, 12), 0.1922704],
 [(2, 3, 12), (2, 3, 10), 0.1760737],
 [(2, 3, 12), (2, 1, 10), 0.0213468],
 [(2, 3, 12), (0, 5, 10), 0.0213468],
 [(2, 3, 12), (0, 3, 10), 0.0427301],
 [(2, 1, 12), (4, 1, 12), 0.1281117],
 [(2, 1, 12), (2, 5, 12), 0.1922704],
 [(2, 1, 12), (2, 5, 10), 0.0213468],
 [(2, 1, 12), (2, 3, 12), 0.1922704],
 [(2, 1, 12), (2, 1, 10), 0.1760737],
 [(2, 1, 12), (0, 3, 10), 0.0213468],
 [(2, 1, 12), (0, 1, 10), 0.0427301],
 [(0, 5, 12), (4, 5, 12), 0.1281117],
 [(0, 5, 12), (4, 5, 10), 0.0427301],
 [(0, 5, 12), (4, 1, 10), 0.0213468],
 [(0, 5, 12), (2, 5, 12), 0.1281117],
 [(0, 5, 12), (0, 5, 10), 0.1760737],
 [(0, 5, 12), (0, 3, 10), 0.0213468],
 [(0, 3, 12), (4, 5, 10), 0.0213468],
 [(0, 3, 12), (4, 3, 12), 0.1281117],
 [(0, 3, 12), (4, 3, 10), 0.0427301],
 [(0, 3, 12), (2, 3, 12), 0.1281117],
 [(0, 3, 12), (0, 5, 12), 0.1922704],
 [(0, 3, 12), (0, 3, 10), 0.1760737],
 [(0, 3, 12), (0, 1, 10), 0.0213468],
 [(0, 1, 12), (4, 3, 10), 0.0213468],
 [(0, 1, 12), (4, 1, 12), 0.1281117],
 [(0, 1, 12), (4, 1, 10), 0.0427301],
 [(0, 1, 12), (2, 1, 12), 0.1281117],
 [(0, 1, 12), (0, 5, 12), 0.1922704],
 [(0, 1, 12), (0, 5, 10), 0.0213468],
 [(0, 1, 12), (0, 3, 12), 0.1922704],
 [(0, 1, 12), (0, 1, 10), 0.1760737],
 [(4, 5, 10), (4, 5, 8), 0.1760737],
 [(4, 5, 10), (4, 3, 8), 0.0213468],
 [(4, 5, 10), (2, 5, 8), 0.0427301],
 [(4, 5, 10), (2, 1, 8), 0.0213468],
 [(4, 3, 10), (4, 5, 10), 0.1922704],
 [(4, 3, 10), (4, 3, 8), 0.1760737],
 [(4, 3, 10), (4, 1, 8), 0.0213468],
 [(4, 3, 10), (2, 5, 8), 0.0213468],
 [(4, 3, 10), (2, 3, 8), 0.0427301],
 [(4, 1, 10), (4, 5, 10), 0.1922704],
 [(4, 1, 10), (4, 5, 8), 0.0213468],
 [(4, 1, 10), (4, 3, 10), 0.1922704],
 [(4, 1, 10), (4, 1, 8), 0.1760737],
 [(4, 1, 10), (2, 3, 8), 0.0213468],
 [(4, 1, 10), (2, 1, 8), 0.0427301],
 [(2, 5, 10), (4, 5, 10), 0.1281117],
 [(2, 5, 10), (2, 5, 8), 0.1760737],
 [(2, 5, 10), (2, 3, 8), 0.0213468],
 [(2, 5, 10), (0, 5, 8), 0.0427301],
 [(2, 5, 10), (0, 1, 8), 0.0213468],
 [(2, 3, 10), (4, 3, 10), 0.1281117],
 [(2, 3, 10), (2, 5, 10), 0.1922704],
 [(2, 3, 10), (2, 3, 8), 0.1760737],
 [(2, 3, 10), (2, 1, 8), 0.0213468],
 [(2, 3, 10), (0, 5, 8), 0.0213468],
 [(2, 3, 10), (0, 3, 8), 0.0427301],
 [(2, 1, 10), (4, 1, 10), 0.1281117],
 [(2, 1, 10), (2, 5, 10), 0.1922704],
 [(2, 1, 10), (2, 5, 8), 0.0213468],
 [(2, 1, 10), (2, 3, 10), 0.1922704],
 [(2, 1, 10), (2, 1, 8), 0.1760737],
 [(2, 1, 10), (0, 3, 8), 0.0213468],
 [(2, 1, 10), (0, 1, 8), 0.0427301],
 [(0, 5, 10), (4, 5, 10), 0.1281117],
 [(0, 5, 10), (4, 5, 8), 0.0427301],
 [(0, 5, 10), (4, 1, 8), 0.0213468],
 [(0, 5, 10), (2, 5, 10), 0.1281117],
 [(0, 5, 10), (0, 5, 8), 0.1760737],
 [(0, 5, 10), (0, 3, 8), 0.0213468],
 [(0, 3, 10), (4, 5, 8), 0.0213468],
 [(0, 3, 10), (4, 3, 10), 0.1281117],
 [(0, 3, 10), (4, 3, 8), 0.0427301],
 [(0, 3, 10), (2, 3, 10), 0.1281117],
 [(0, 3, 10), (0, 5, 10), 0.1922704],
 [(0, 3, 10), (0, 3, 8), 0.1760737],
 [(0, 3, 10), (0, 1, 8), 0.0213468],
 [(0, 1, 10), (4, 3, 8), 0.0213468],
 [(0, 1, 10), (4, 1, 10), 0.1281117],
 [(0, 1, 10), (4, 1, 8), 0.0427301],
 [(0, 1, 10), (2, 1, 10), 0.1281117],
 [(0, 1, 10), (0, 5, 10), 0.1922704],
 [(0, 1, 10), (0, 5, 8), 0.0213468],
 [(0, 1, 10), (0, 3, 10), 0.1922704],
 [(0, 1, 10), (0, 1, 8), 0.1760737],
 [(4, 5, 8), (4, 5, 6), 0.1760737],
 [(4, 5, 8), (4, 3, 6), 0.0213468],
 [(4, 5, 8), (2, 5, 6), 0.0427301],
 [(4, 5, 8), (2, 1, 6), 0.0213468],
 [(4, 3, 8), (4, 5, 8), 0.1922704],
 [(4, 3, 8), (4, 3, 6), 0.1760737],
 [(4, 3, 8), (4, 1, 6), 0.0213468],
 [(4, 3, 8), (2, 5, 6), 0.0213468],
 [(4, 3, 8), (2, 3, 6), 0.0427301],
 [(4, 1, 8), (4, 5, 8), 0.1922704],
 [(4, 1, 8), (4, 5, 6), 0.0213468],
 [(4, 1, 8), (4, 3, 8), 0.1922704],
 [(4, 1, 8), (4, 1, 6), 0.1760737],
 [(4, 1, 8), (2, 3, 6), 0.0213468],
 [(4, 1, 8), (2, 1, 6), 0.0427301],
 [(2, 5, 8), (4, 5, 8), 0.1281117],
 [(2, 5, 8), (2, 5, 6), 0.1760737],
 [(2, 5, 8), (2, 3, 6), 0.0213468],
 [(2, 5, 8), (0, 5, 6), 0.0427301],
 [(2, 5, 8), (0, 1, 6), 0.0213468],
 [(2, 3, 8), (4, 3, 8), 0.1281117],
 [(2, 3, 8), (2, 5, 8), 0.1922704],
 [(2, 3, 8), (2, 3, 6), 0.1760737],
 [(2, 3, 8), (2, 1, 6), 0.0213468],
 [(2, 3, 8), (0, 5, 6), 0.0213468],
 [(2, 3, 8), (0, 3, 6), 0.0427301],
 [(2, 1, 8), (4, 1, 8), 0.1281117],
 [(2, 1, 8), (2, 5, 8), 0.1922704],
 [(2, 1, 8), (2, 5, 6), 0.0213468],
 [(2, 1, 8), (2, 3, 8), 0.1922704],
 [(2, 1, 8), (2, 1, 6), 0.1760737],
 [(2, 1, 8), (0, 3, 6), 0.0213468],
 [(2, 1, 8), (0, 1, 6), 0.0427301],
 [(0, 5, 8), (4, 5, 8), 0.1281117],
 [(0, 5, 8), (4, 5, 6), 0.0427301],
 [(0, 5, 8), (4, 1, 6), 0.0213468],
 [(0, 5, 8), (2, 5, 8), 0.1281117],
 [(0, 5, 8), (0, 5, 6), 0.1760737],
 [(0, 5, 8), (0, 3, 6), 0.0213468],
 [(0, 3, 8), (4, 5, 6), 0.0213468],
 [(0, 3, 8), (4, 3, 8), 0.1281117],
 [(0, 3, 8), (4, 3, 6), 0.0427301],
 [(0, 3, 8), (2, 3, 8), 0.1281117],
 [(0, 3, 8), (0, 5, 8), 0.1922704],
 [(0, 3, 8), (0, 3, 6), 0.1760737],
 [(0, 3, 8), (0, 1, 6), 0.0213468],
 [(0, 1, 8), (4, 3, 6), 0.0213468],
 [(0, 1, 8), (4, 1, 8), 0.1281117],
 [(0, 1, 8), (4, 1, 6), 0.0427301],
 [(0, 1, 8), (2, 1, 8), 0.1281117],
 [(0, 1, 8), (0, 5, 8), 0.1922704],
 [(0, 1, 8), (0, 5, 6), 0.0213468],
 [(0, 1, 8), (0, 3, 8), 0.1922704],
 [(0, 1, 8), (0, 1, 6), 0.1760737],
 [(4, 5, 6), (4, 5, 4), 0.1760737],
 [(4, 5, 6), (4, 3, 4), 0.0213468],
 [(4, 5, 6), (2, 5, 4), 0.0427301],
 [(4, 5, 6), (2, 1, 4), 0.0213468],
 [(4, 3, 6), (4, 5, 6), 0.1922704],
 [(4, 3, 6), (4, 3, 4), 0.1760737],
 [(4, 3, 6), (4, 1, 4), 0.0213468],
 [(4, 3, 6), (2, 5, 4), 0.0213468],
 [(4, 3, 6), (2, 3, 4), 0.0427301],
 [(4, 1, 6), (4, 5, 6), 0.1922704],
 [(4, 1, 6), (4, 5, 4), 0.0213468],
 [(4, 1, 6), (4, 3, 6), 0.1922704], *
 [(4, 1, 6), (4, 1, 4), 0.1760737],
 [(4, 1, 6), (2, 3, 4), 0.0213468],
 [(4, 1, 6), (2, 1, 4), 0.0427301],
 [(2, 5, 6), (4, 5, 6), 0.1281117],
 [(2, 5, 6), (2, 5, 4), 0.1760737],
 [(2, 5, 6), (2, 3, 4), 0.0213468],
 [(2, 5, 6), (0, 5, 4), 0.0427301],
 [(2, 5, 6), (0, 1, 4), 0.0213468],
 [(2, 3, 6), (4, 3, 6), 0.1281117],
 [(2, 3, 6), (2, 5, 6), 0.1922704],
 [(2, 3, 6), (2, 3, 4), 0.1760737],
 [(2, 3, 6), (2, 1, 4), 0.0213468],
 [(2, 3, 6), (0, 5, 4), 0.0213468],
 [(2, 3, 6), (0, 3, 4), 0.0427301],
 [(2, 1, 6), (4, 1, 6), 0.1281117],
 [(2, 1, 6), (2, 5, 6), 0.1922704],
 [(2, 1, 6), (2, 5, 4), 0.0213468],
 [(2, 1, 6), (2, 3, 6), 0.1922704],
 [(2, 1, 6), (2, 1, 4), 0.1760737],
 [(2, 1, 6), (0, 3, 4), 0.0213468],
 [(2, 1, 6), (0, 1, 4), 0.0427301],
 [(0, 5, 6), (4, 5, 6), 0.1281117],
 [(0, 5, 6), (4, 5, 4), 0.0427301],
 [(0, 5, 6), (4, 1, 4), 0.0213468],
 [(0, 5, 6), (2, 5, 6), 0.1281117],
 [(0, 5, 6), (0, 5, 4), 0.1760737],
 [(0, 5, 6), (0, 3, 4), 0.0213468],
 [(0, 3, 6), (4, 5, 4), 0.0213468],
 [(0, 3, 6), (4, 3, 6), 0.1281117],
 [(0, 3, 6), (4, 3, 4), 0.0427301],
 [(0, 3, 6), (2, 3, 6), 0.1281117],
 [(0, 3, 6), (0, 5, 6), 0.1922704],
 [(0, 3, 6), (0, 3, 4), 0.1760737],
 [(0, 3, 6), (0, 1, 4), 0.0213468],
 [(0, 1, 6), (4, 3, 4), 0.0213468],
 [(0, 1, 6), (4, 1, 6), 0.1281117],
 [(0, 1, 6), (4, 1, 4), 0.0427301],
 [(0, 1, 6), (2, 1, 6), 0.1281117],
 [(0, 1, 6), (0, 5, 6), 0.1922704],
 [(0, 1, 6), (0, 5, 4), 0.0213468],
 [(0, 1, 6), (0, 3, 6), 0.1922704],
 [(0, 1, 6), (0, 1, 4), 0.1760737],
 [(4, 5, 4), (4, 5, 2), 0.1760737],
 [(4, 5, 4), (4, 3, 2), 0.0213468],
 [(4, 5, 4), (2, 5, 2), 0.0427301],
 [(4, 5, 4), (2, 1, 2), 0.0213468],
 [(4, 3, 4), (4, 5, 4), 0.1922704],
 [(4, 3, 4), (4, 3, 2), 0.1760737],
 [(4, 3, 4), (4, 1, 2), 0.0213468],
 [(4, 3, 4), (2, 5, 2), 0.0213468],
 [(4, 3, 4), (2, 3, 2), 0.0427301],
 [(4, 1, 4), (4, 5, 4), 0.1922704],
 [(4, 1, 4), (4, 5, 2), 0.0213468],
 [(4, 1, 4), (4, 3, 4), 0.1922704],
 [(4, 1, 4), (4, 1, 2), 0.1760737],
 [(4, 1, 4), (2, 3, 2), 0.0213468],
 [(4, 1, 4), (2, 1, 2), 0.0427301],
 [(2, 5, 4), (4, 5, 4), 0.1281117],
 [(2, 5, 4), (2, 5, 2), 0.1760737],
 [(2, 5, 4), (2, 3, 2), 0.0213468],
 [(2, 5, 4), (0, 5, 2), 0.0427301],
 [(2, 5, 4), (0, 1, 2), 0.0213468],
 [(2, 3, 4), (4, 3, 4), 0.1281117],
 [(2, 3, 4), (2, 5, 4), 0.1922704],
 [(2, 3, 4), (2, 3, 2), 0.1760737],
 [(2, 3, 4), (2, 1, 2), 0.0213468],
 [(2, 3, 4), (0, 5, 2), 0.0213468],
 [(2, 3, 4), (0, 3, 2), 0.0427301],
 [(2, 1, 4), (4, 1, 4), 0.1281117],
 [(2, 1, 4), (2, 5, 4), 0.1922704],
 [(2, 1, 4), (2, 5, 2), 0.0213468],
 [(2, 1, 4), (2, 3, 4), 0.1922704],
 [(2, 1, 4), (2, 1, 2), 0.1760737],
 [(2, 1, 4), (0, 3, 2), 0.0213468],
 [(2, 1, 4), (0, 1, 2), 0.0427301],
 [(0, 5, 4), (4, 5, 4), 0.1281117],
 [(0, 5, 4), (4, 5, 2), 0.0427301],
 [(0, 5, 4), (4, 1, 2), 0.0213468],
 [(0, 5, 4), (2, 5, 4), 0.1281117],
 [(0, 5, 4), (0, 5, 2), 0.1760737],
 [(0, 5, 4), (0, 3, 2), 0.0213468],
 [(0, 3, 4), (4, 5, 2), 0.0213468],
 [(0, 3, 4), (4, 3, 4), 0.1281117],
 [(0, 3, 4), (4, 3, 2), 0.0427301],
 [(0, 3, 4), (2, 3, 4), 0.1281117],
 [(0, 3, 4), (0, 5, 4), 0.1922704],
 [(0, 3, 4), (0, 3, 2), 0.1760737],
 [(0, 3, 4), (0, 1, 2), 0.0213468],
 [(0, 1, 4), (4, 3, 2), 0.0213468],
 [(0, 1, 4), (4, 1, 4), 0.1281117],
 [(0, 1, 4), (4, 1, 2), 0.0427301],
 [(0, 1, 4), (2, 1, 4), 0.1281117],
 [(0, 1, 4), (0, 5, 4), 0.1922704],
 [(0, 1, 4), (0, 5, 2), 0.0213468],
 [(0, 1, 4), (0, 3, 4), 0.1922704],
 [(0, 1, 4), (0, 1, 2), 0.1760737],
 [(4, 5, 2), (4, 5, 0), 0.1760737],
 [(4, 5, 2), (4, 3, 0), 0.0213468],
 [(4, 5, 2), (2, 5, 0), 0.0427301],
 [(4, 5, 2), (2, 1, 0), 0.0213468],
 [(4, 3, 2), (4, 5, 2), 0.1922704],
 [(4, 3, 2), (4, 3, 0), 0.1760737],
 [(4, 3, 2), (4, 1, 0), 0.0213468],
 [(4, 3, 2), (2, 5, 0), 0.0213468],
 [(4, 3, 2), (2, 3, 0), 0.0427301],
 [(4, 1, 2), (4, 5, 2), 0.1922704],
 [(4, 1, 2), (4, 5, 0), 0.0213468],
 [(4, 1, 2), (4, 3, 2), 0.1922704],
 [(4, 1, 2), (4, 1, 0), 0.1760737],
 [(4, 1, 2), (2, 3, 0), 0.0213468],
 [(4, 1, 2), (2, 1, 0), 0.0427301],
 [(2, 5, 2), (4, 5, 2), 0.1281117],
 [(2, 5, 2), (2, 5, 0), 0.1760737],
 [(2, 5, 2), (2, 3, 0), 0.0213468],
 [(2, 5, 2), (0, 5, 0), 0.0427301],
 [(2, 5, 2), (0, 1, 0), 0.0213468],
 [(2, 3, 2), (4, 3, 2), 0.1281117],
 [(2, 3, 2), (2, 5, 2), 0.1922704],
 [(2, 3, 2), (2, 3, 0), 0.1760737],
 [(2, 3, 2), (2, 1, 0), 0.0213468],
 [(2, 3, 2), (0, 5, 0), 0.0213468],
 [(2, 3, 2), (0, 3, 0), 0.0427301],
 [(2, 1, 2), (4, 1, 2), 0.1281117],
 [(2, 1, 2), (2, 5, 2), 0.1922704],
 [(2, 1, 2), (2, 5, 0), 0.0213468],
 [(2, 1, 2), (2, 3, 2), 0.1922704],
 [(2, 1, 2), (2, 1, 0), 0.1760737],
 [(2, 1, 2), (0, 3, 0), 0.0213468],
 [(2, 1, 2), (0, 1, 0), 0.0427301],
 [(0, 5, 2), (4, 5, 2), 0.1281117],
 [(0, 5, 2), (4, 5, 0), 0.0427301],
 [(0, 5, 2), (4, 1, 0), 0.0213468],
 [(0, 5, 2), (2, 5, 2), 0.1281117],
 [(0, 5, 2), (0, 5, 0), 0.1760737],
 [(0, 5, 2), (0, 3, 0), 0.0213468],
 [(0, 3, 2), (4, 5, 0), 0.0213468],
 [(0, 3, 2), (4, 3, 2), 0.1281117],
 [(0, 3, 2), (4, 3, 0), 0.0427301],
 [(0, 3, 2), (2, 3, 2), 0.1281117],
 [(0, 3, 2), (0, 5, 2), 0.1922704],
 [(0, 3, 2), (0, 3, 0), 0.1760737],
 [(0, 3, 2), (0, 1, 0), 0.0213468],
 [(0, 1, 2), (4, 3, 0), 0.0213468],
 [(0, 1, 2), (4, 1, 2), 0.1281117],
 [(0, 1, 2), (4, 1, 0), 0.0427301],
 [(0, 1, 2), (2, 1, 2), 0.1281117],
 [(0, 1, 2), (0, 5, 2), 0.1922704],
 [(0, 1, 2), (0, 5, 0), 0.0213468],
 [(0, 1, 2), (0, 3, 2), 0.1922704],
 [(0, 1, 2), (0, 1, 0), 0.1760737],
 [(4, 3, 0), (4, 5, 0), 0.0960532],
 [(4, 1, 0), (4, 5, 0), 0.0960532],
 [(4, 1, 0), (4, 3, 0), 0.0960532],
 [(2, 5, 0), (4, 5, 0), 0.0640012],
 [(2, 3, 0), (4, 3, 0), 0.0640012],
 [(2, 3, 0), (2, 5, 0), 0.0960532],
 [(2, 1, 0), (4, 1, 0), 0.0640012],
 [(2, 1, 0), (2, 5, 0), 0.0960532],
 [(2, 1, 0), (2, 3, 0), 0.0960532],
 [(0, 5, 0), (4, 5, 0), 0.0640012],
 [(0, 5, 0), (2, 5, 0), 0.0640012],
 [(0, 3, 0), (4, 3, 0), 0.0640012],
 [(0, 3, 0), (2, 3, 0), 0.0640012],
 [(0, 3, 0), (0, 5, 0), 0.0960532],
 [(0, 1, 0), (4, 1, 0), 0.0640012],
 [(0, 1, 0), (2, 1, 0), 0.0640012],
 [(0, 1, 0), (0, 5, 0), 0.0960532],
 [(0, 1, 0), (0, 3, 0), 0.0960532]
]
\end{verbatim}

\end{document}